\documentclass[aps,prl,preprint]{revtex4}
\usepackage{epsfig}

\begin{document}

\title{MITOCHONDRIAL DNA REPLACEMENT VERSUS
NUCLEAR DNA PERSISTENCE}

\author{Maurizio Serva}  
\affiliation{Dipartimento di Matematica,
Universit\`a dell'Aquila,
I-67010 L'Aquila, Italy}

\bigskip

\date{\today}

\begin{abstract}  

In this paper we consider two populations whose
generations are not overlapping and whose size is large. 
The number of males and females in both populations is constant.
Any generation is replaced by a new one and any individual has two
parents for what concerns nuclear DNA and a single one (the mother)
for what concerns mtDNA.
Moreover, at any generation some individuals migrate from the 
first population to the second.

In a finite random time $T$, the mtDNA of the second
population is completely replaced by the mtDNA of the first. 
In the same time,  the nuclear DNA is not completely replaced and a 
fraction $F$ of the ancient nuclear DNA persists.
We compute both $T$ and $F$.
Since this study shows that complete replacement of mtDNA in a
population is compatible with the persistence of a large fraction of
nuclear DNA, it may have some relevance for the Out of
Africa/Multiregional debate in Paleoanthropology.

\noindent
Pacs: 87.23.Kg, 05.40.-a
\end{abstract}

\maketitle

\section{Introduction}

Mitochondrial DNA (mtDNA) is inherited in a
haploid manner through females.
Since its mutation rate is high and can be easily measured, 
mtDNA is a powerful tool for tracking matrilineages 
and it has been widely used in this role by molecular 
biologists.
On the contrary, nuclear DNA is inherited by both parents
and it recombines at any generation.
We show in this paper that
haploid reproduction allows for a complete replacement
of the mtDNA of a population by the mtDNA of immigrants.
On the contrary, diploid reproduction allows for some 
of the ancient nuclear DNA to persist.

We consider two interbreeding populations 
whose generations are not overlapping and whose size is large 
and constant in time. 
The number of males and females is the same and it is constant
both in the first and in the second population.
Any generation is replaced by a new one and any individual has two
parents for what concerns nuclear DNA and a single one (the mother)
for what concerns mtDNA.
One of the two populations, that we call the African population,
produces some emigrants at any generation ($2p$ on average). 
The second population, that we call the
Asian population, receives these people as immigrants.
The size of the two populations is not necessarily the same
and we assume that the number of African females is $M$ while the 
number of Asian females is $N$, so that the total number of
individuals in the two populations is $2N+2M$.

Let us now explain how reproduction and migration are modeled.
We assume that any individual in the new 
generation chooses independently the two parents at random 
in the previous one
(see~\cite{SP,DJM,DMZ}).
The choice of an individual is independent on the choice of the others.
Moreover, the Africans always choose among Africans
while Asians choose with probability $1-p/N$ among Asians
and with probability $p/N$ among Africans.
The choice is neutral, i.e. there is not preferred choice among
African individuals as well there is not 
a preferred choice among Asian ones. 
The choice of an African parent with probability $p/N$
is equivalent to a migration of $2p$ Africans on average,
one half of which, still on average, are females.
Remark that the number of emigrants remains finite even if the
population becomes very large ($N$  $\to$  $\infty$).

If the migration rate $p$ vanishes,
the mtDNA of a population can not be transmitted to the other.
In this case,  both the African mtDNA and the Asian one separately 
undergo to standard coalescence 
(see~\cite{K1,K2} and more recently~\cite{S,SD}
for dynamical aspects).
On the contrary, if migration is allowed, 
we show that in a finite random time $T$
the mtDNA of the Asian population is completely replaced
by the mtDNA of immigrants 
(time is the number of generations divided by $N$). 
We also show that, in the same time, a fraction $F$ of the
nuclear DNA persists in the Asian population.
In other words, the mtDNA of the Asian population living 
at a time $T$ before present completely disappeared in the present
population while a fraction $F$ of the ancient nuclear DNA
still persists
(hereafter the  'ancient nuclear DNA' is the nuclear DNA of
the Asian population living $TN$ generations before present).
Complete mtDNA replacement together with nuclear DNA persistence
occur even if the migration rate is very low.
\smallskip

We find the random replacement time $T$ in the second section
and the fraction $F$ of ancient nuclear DNA in the third.
In the last section we discuss the eventual relevance 
of our results for the debate about Out of Africa
and Multiregional  models in Paleoanthropology.
The mathematical core of the paper is the Appendix
where we show that the fraction of ancient nuclear DNA is 
a deterministic quantity which decreases exponentially in time.
This is the reason why the ancient nuclear DNA 
can be diluted by the nuclear DNA of immigrants 
but it cannot be totally replaced.

\section{Replacement Time}

Let us start with the following remark:
since Africans always choose among Africans,
African mtDNA undergoes to standard coalescence.
The average coalescence time for African females 
is $2M/N$ which means $2M$ generations
(the probability density for coalescence time 
can be found, for example, in ~\cite{S}). 
On the contrary, present Asian female population may have mtDNA 
ancestors both in the Asian and in the African population.
Assume that at a given time in the past the number of Asian mtDNA
female ancestors is $n$. Going backward for one generation, 
the probability that this number decreases to $n-1$ is $\frac{b(n)}{N}$ 
where $b(n)=pn+n(n-1)/2$.
The term $\frac{pn}{N}$
is due to the probability that one of the female ancestors choose
an  African mother and the term $\frac{n(n-1)/2}{N}$ is due
the probability that two of the female ancestors choose
the same Asian mother (the celebrated coalescence phenomenon).
Than, the probability that the number of ancestors remains the same 
going backward for $tN$ generations is  
$[1-\frac{b(n)}{N}]^{tN}$ which, for large $N$, becomes
$\exp(-b(n)t)$.
Therefore, the time $t_n$ needed
for reducing from $n$ to $n-1$ the number
of Asian female ancestors is exponentially
distributed with average $[pn+n(n-1)/2]^{-1}$.
Finally, the time needed for complete replacement, i.e. for
the present Asian population to have not Asian mtDNA ancestors, is 
 
\begin{equation}
T= \sum_{j=1}^\infty t_j
\label{time}
\end{equation}
which is a sum of independent random times
exponentially distributed and with parameters $[b(j)]^{-1}$.
Remark that the above sum starts from one 
(not from two as in the coalescent) since
complete replacement occurs when the
number of Asian ancestors vanishes.

The average replacement time is than 

\begin{equation} 
<T>= \sum_{j=1}^\infty [jp+ \frac{j(j-1)}{2}]^{-1}
\label{average}
\end{equation}
which is plotted in the figure.

It should be remarked that the number of emigrants
which is requested for the rapid replacement of mtDNA is very small.
A couple of immigrants at any generation ($p=1$) in a large population of 
size $N$ allows for a complete replacement of the Asian mtDNA in
about $2N$ generations, while, if the 
couples are two ($p=2$), it is sufficient $60\%$ of the time.

\begin{figure}
\vspace{.2in}
\centerline{\psfig{figure=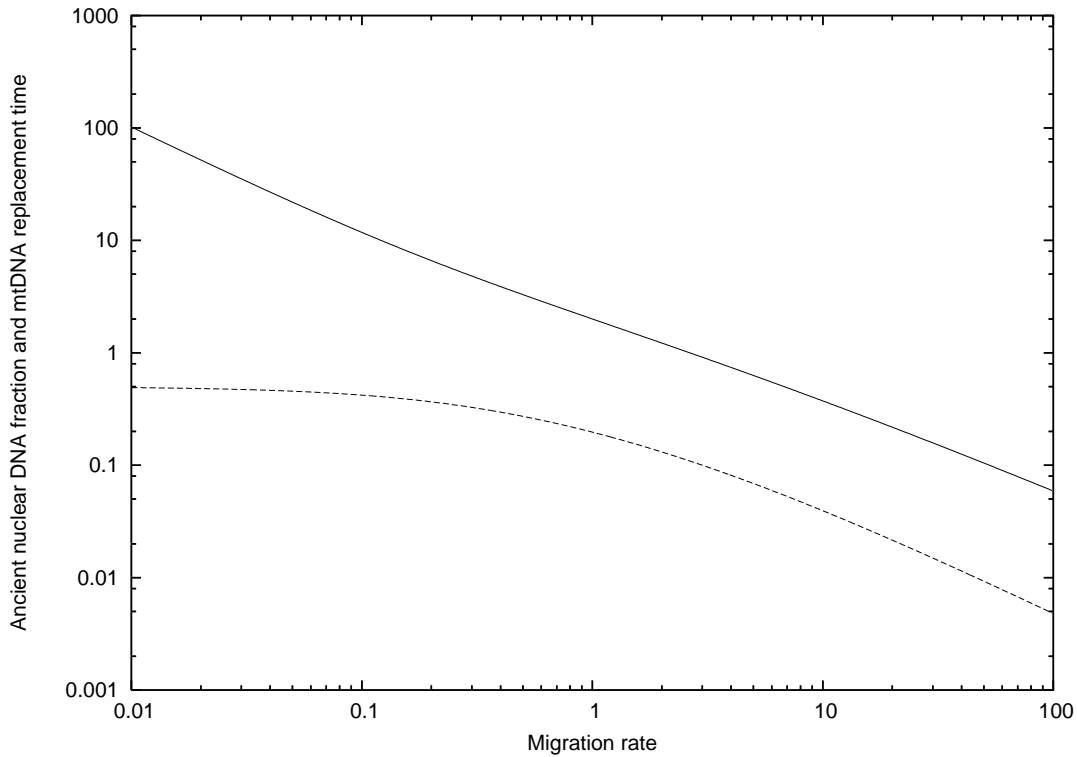,width=4.0truein,angle=270}}
\bigskip
\caption{
Average of mtDNA replacement time (full line)
and average of ancient nuclear DNA fraction (dashed line)
versus emigration rate $p$.}
\label{f1}
\end{figure}

\section{Ancient Nuclear DNA}

We have shown that the mtDNA of the Asian generation 
at a time $T$ in the past (the ancient generation)
completely disappeared in present Asian population.
We want to see now what happened to the nuclear DNA of 
that generation (the ancient DNA).

Let us first remark that the fraction of ancient 
DNA for the ancient generation equals 1, while,
at any following generation, this fraction is less than 1.
In fact, at any generation replacement, both the father and the mother
are independently chosen with probability $p/N$ among Africans
and with probability $1-p/N$ among Asians.
Since nuclear DNA of an individual
comes for one half from the father and for one half from the mother,
at any generation replacement, the fraction
of ancient nuclear DNA in the entire Asian population 
is reduced, on average, by a factor $(1-\frac{p}{N})$.
The non averaged factor, indeed, randomly fluctuates around this value.
Fluctuations are due to the fact that the number of
immigrants is random and are also due to the
Wright-Fisher diffusion associated with generation replacement.
Fluctuations of both origins are of order $1/N$
as it is shown in the Appendix.

Fluctuations are small because diploid reproduction
is able to rapidly span ancient and new genes in all population.
This self-averaging behavior is at very variance with
haploid reproduction where a single individual
only may have new mtDNA  or ancient one.  

After a time $T$ from the ancient generation (i.e. in the present
population), the average fraction of ancient nuclear DNA
reduces to  $(1-\frac{p}{N})^{TN}$, which, for large $N$ becomes 
$\exp(-pT)$.

For each generation replacement, random fluctuations are of 
the order $1/N$ and, since they are uncorrelated,
after $TN$ generations they are of order $1/\sqrt(N)$
(see the Appendix).
Therefore, the fraction of ancient DNA 
in present generation is exactly $F=\exp{(-pT)}$ when the population
is large and randomness is only due to the randomness of the time $T$.
It must be remarked that $F$ is never vanishing,
since the random time $T$ is finite with probability one.
This implies that nuclear DNA is never completely replaced, 
at variance with mtDNA.

As already remarked, this self-averaging behavior in the large $N$ limit
is typical of diploid reproduction (see~\cite{SP,DJM,DMZ}).
Indeed, self-averaging, which is proved in the Appendix,
is the key point of our results since 
the persistence of a fraction of ancient nuclear DNA, 
when mtDNA is totally replaced, is a 
direct consequence of it.

In order to see how $F$ depends on $p$
we take now the average of $F$ with respect to $T$ and 
we obtain

\begin{equation} 
<F>=<\exp{(-pT)}>= \prod_{n=1}^\infty \frac{2np+n(n-1)}{2(n+1)p+n(n-1)}
\label{averagefraction}
\end{equation}
which is also plotted in the figure.

As already remarked, $p=1$ allows for a complete replacement 
of the Asian mtDNA in about $2N$ generations, while for $p=2$ 
a number $1.2 \, N$ of generations is sufficient.
In the first case the average fraction of ancient nuclear DNA is
$0.2$ while in the second is $0.12$.
The point is to compare the average replacement time $<T>$
with African average coalescence time $2M/N$.
Assuming that African population size is equal or larger than 
Asian population size ($2M \ge 2N$) one finds that in both the 
above examples $<T>$ is equal or smaller than  $2M/N$.
In other words, one or two couples of immigrants at any generation,
allow for a replacement of Asian mtDNA in a time 
smaller than African coalescence time.
Furthermore, they allow for the persistence of a significant 
fraction of ancient nuclear DNA.
 
\section{Discussion}

Let us now discuss some possible consequences 
concerning Paleoanthropology.
Assume that $N=M=5000$ (number of African and Asian women),
in this case, the average number of generations
requested for coalescence of
African mtDNA is $2M=10,000$ which, assumed generations
of $20$ years, correspond to $200,000$ years.
A couple of immigrants at any generation ($p=1$) induces complete
replacement of the Asian mtDNA in the same time, while, 
for two couples ($p=2$), the time
requested is $120,000$ years.
As already discussed, the fraction of ancient nuclear DNA 
ranges between $0.12$ and $0.2$.
Moreover, if $N$ is smaller than $M=5000$,
the migration rate $p$ requested for having  $<T> \, \le 2M/N$
can be smaller than $1$ and, therefore,  the fraction of 
ancient nuclear DNA can be higher, up to a maximum of $0.5$.

In conclusion, mtDNA argument cannot be used to
prove 'Out of Africa' theory (see~\cite{CW} for a review)
or to disprove  Multiregional Model (see~\cite{TW} for a review)
since a very small migration flux  
is compatible both with pre-African nuclear DNA persistence
and complete pre-African mtDNA replacement in Asia and Europe.
Indeed, the picture in this paper is compatible
with~\cite{T}, where a study of worldwide
human nuclear DNA seems to show repeated migrations form Africa to Europe and
Asia.

Finally, we would like to mention, that y-chromosome
is also inherited in an haploid manner,
the only difference is that its reproduction is driven by males. 
The qualitative and quantitative arguments in the paper remain
unchanged if y-chromosome is considered in place of mtDNA.

\section{Acknowledgments}

We thank Davide Gabrielli and Filippo Petroni
for many discussions and for 
critical reading of the manuscript.
We also acknowledge the financial support of the MIUR - Universit\`a di
L'Aquila Cofin 2004 n. 2004028108-005.

\section{Appendix}

Let us define $m_i(t)$ as the fraction of ancient nuclear DNA
for the male individual $i$ at $tN$ generations 
after the ancient generation
and let also define the analogous $f_i(t)$ for the female individual $i$.
By definition the fraction of ancient DNA
for an individual of the ancient generation equals one,
than we have $m_i(0)=f_i(0)=1$.
In the reproductive process an individual receives 
the nuclear DNA of both parents
so that his fraction of ancient nuclear DNA will be an average
of the fraction of the two parents.
Indeed, this hold exactly only for large genomes 
as the human one. 
The link between two generations is than provided
by the following stochastic equations 

\begin{equation} 
m_i(t)= \frac{1}{2}
[\mu_i(t) \, m_{j(i,t)}(t-\epsilon) +\phi_i(t) \, f_{k(i,t)}(t-\epsilon)]
\label{4}
\end{equation}
Where $\epsilon \equiv 1/N$.
The variables $j(i,t)$ and $k(i,t)$ take 
any integer value between 1 and $N$
with equal probability $1/N$. 
The variables $\mu_i(t)$ and  $\phi_i(t)$ take 
the values $0$ with probability $p/N$ and 1 
with probability  $1-p/N$. 
With our choice the father contribution $\frac{1}{2}
\mu_i(t) \, m_{j(i,t)}(t-\epsilon)$ to the
fraction $m_i(t)$ vanishes with probability 
$p/N$ (African father) and equals one half of the 
fraction of a given Asian father with probability $(1-p/N)/N$.
The same for the mother contribution.
All variables $\mu_i(t)$,  $\phi_i(t)$, $j(i,t)$, and $k(i,t)$,
are mutually independent,  furthermore, two variables of same type
are independent  whenever the 
individual indexes or time indexes are different
(for example $j(i,t)$ and $j(k,s)$ are independent 
when $i \neq j$ and/or $t \neq s$). 

An analogous stochastic equations  holds for $f_i(t)$
with all variables in it independent from
the mirror variables in (\ref{4}).

Since $<\mu_i(t)>$ = $<\phi_i(t)>$ = $(1-p/N)$
and since both $j(i,t)$ and $k(i,t)$ take 
any integer value between 1 and $N$
with equal probability $1/N$, we obtain
by averaging (\ref{4})

\begin{equation} 
<m_i(t)>= \frac{1}{2N} (1-\frac{p}{N})
\sum_{l=1}^N ( <m_l(t-\epsilon)> + <f_l(t-\epsilon)> ) 
\label{5}
\end{equation}

Since averages are independent on the individual index,
and averages for females and males must coincide
($<m_i(t)>$ = $<f_i(t)>$),
we have

\begin{equation} 
<m_i(t)>= (1-\frac{p}{N})<m_i(t-\epsilon)>=(1-\frac{p}{N})^{tN}
\label{6}
\end{equation}
where the second equality is obtained by iteration and by 
the initial condition $m_i(0)=1$.

Analogously,
$<[\mu_i(t)]^2>$ =$<[\phi_i(t)]^2>$ = $(1-p/N)$ 
and 
$<m_i(t) f_i(t)>$ =$<m_i(t) f_j(t)>$ = $<m_i(t) m_j(t)>$
when $i \neq j$
as it can be easily verified using (\ref{4}).
Than we have

\begin{equation} 
<[m_i(t)]^2>= \frac{1}{2} (1-\frac{p}{N})
<[m_i(t-\epsilon)]^2> 
+ \frac{1}{2} (1-\frac{p}{N})^2
<m_i(t-\epsilon) \, m_j(t-\epsilon)>
\label{7}
\end{equation}
where it is intended that  $i \neq j$ and were
we have again made use of the independence
of averages on the individual index,
and of the fact that averages for females and males coincide
($<[m_i(t)]^2>$ = $<[f_i(t)]^2>$)

Since $\epsilon = 1/N$, this equality can hold only if

\begin{equation} 
<[m_i(t)]^2>= 
<m_i(t) \, m_j(t)>+ o(\frac{1}{N})
\label{8}
\end{equation}
where  $o(\frac{1}{N})$ means 'of order $1/N$'.

This is the key point since we will use it
to prove that ancient nuclear DNA fraction behaves deterministically
in large populations.
It is important to remark that the above 
equality is associated to diploid reproduction.
In fact, (\ref{8}) is a direct consequence of
the fact that nuclear DNA
is an average of that of both parents
as described by equation (\ref{4}).
For haploid reproduction  an analogous of equation (\ref{4}) holds,
but the contribution comes only from a single parent
and equality (\ref{8}) cannot be stated.

Let us now define the fraction of ancient nuclear DNA in a population
as the mean of the fraction of the component individuals

\begin{equation} 
x(t)= \frac{1}{2N}\sum_{i=1}^N [m_i(t) +f_i(t)]
\label{9}
\end{equation}
then, from (\ref{6}) we immediately obtain

\begin{equation} 
<x(t)>=(1-\frac{p}{N})^{tN}
\label{10}
\end{equation}

We will show now that $x(t)$ is a deterministic variable
in large population and, therefore, coincides with its average.
Using again individual index symmetries and male/female symmetry
we obtain from (\ref{4}) and (\ref{9})

\begin{equation} 
<[x(t)]^2>=(1-\frac{p}{N})^2 \,< [x(t-\epsilon)]^2>
+ R(t-\epsilon)
\label{11}
\end{equation}
where

\begin{equation} 
R(t)=\frac{1}{2N}(1-\frac{p}{N})
(<[m_i(t)]^2>-(1-\frac{p}{N})<m_i(t) \, m_j(t)>)
\label{12}
\end{equation}

Using equality (\ref{8}) we obtain  $R(t)=o(\frac{1}{N^2})$.
Indeed this is the key of the proof, 
in fact, for haploid DNA transmission,
we would obtain the same equation (\ref{11}) 
but we would have
$R(t)=o(\frac{1}{N})$ in place of $R(t)=o(\frac{1}{N^2})$.

From equation  (\ref{11}) together with the condition
$R(t)=o(\frac{1}{N^2})$ we obtain

\begin{equation} 
<[x(t)]^2>=(1-\frac{p}{N})^{2tN} +o(\frac{1}{N})
\label{13}
\end{equation}
which compared with (\ref{10})
tells us that

\begin{equation} 
x(t)= (1-\frac{p}{N})^{tN} \pm o(\frac{1}{\sqrt(N)}) 
\label{14}
\end{equation}

We remark that for haploid reproduction, fluctuations are of order 1
($o(1)$ in place of $o(\frac{1}{\sqrt(N)}$),
that is why mtDNA disappears in a finite random time, even for
large populations.

Finally, in the large $N$ limit, we obtain  
from (\ref{13})  $\, x(t)=\exp{(-pt)}$,
which tells us that ancient nuclear DNA fraction decreases
exponentially. Moreover, since the random 
replacement time is finite with probability one, 
the fraction $F \equiv x(T)=\exp{(-pT)}$ is always finite.


\begin{thebibliography}{99}

\bibitem{CW}
R. L. Cann and A. C. Wilson,
{\it The Recent African Genesis of Humans}, 
Scientific American {\bf 13}, (2003), 54-61

\bibitem{DJM} 
B. Derrida and B. Jung-Muller, 
{\it The genealogical tree of a chromosome}, 
Journal of  Statistical  Physics, {\bf 94}, (1999), 277-298.

\bibitem{DMZ} 
B. Derrida,  S. C. Manrubia and D. H. Zanette,
{\it Statistical properties of genealogical trees}, 
Physical  Review Letters, {\bf 82}, (1999), 1987-1990.

\bibitem{K1} 
J. F. C. Kingman, 
{\it The Coalescent}, 
Stochastic Processes and their Applications, 
{\bf 13}, (1982), 235-248.

\bibitem{K2}
J. F. C. Kingman, 
{\it On the genealogy of large populations.
Essays in statistical science}, 
Journal of Applied Probability,  {\bf 19A}, (1982), 27-43.

\bibitem{S}
M. Serva,
{\it On the genealogy of populations: trees, branches and offspring.}
Journal of Statistical Mechanics: 
theory and experiment, (2005), P07011.

\bibitem{SP}
M. Serva and L. Peliti, 
{\it A statistical model 
of an evolving population with sexual reproduction}, 
Journal of Physics A: Mathematical and General {\bf 24}, 
(1991), L705-L709.

\bibitem{SD}
D. Simon and B. Derrida, 
{\it Evolution of the most recent common ancestor of a population 
with no selection}, 
Journal of Statistical Mechanics: 
theory and experiment, (2006), P05002.

\bibitem{T}
A. Templeton, 
{\it Out of Africa: again and again}, 
Nature {\bf 416}, (2002), 45-51.


\bibitem{TW}
A. G. Thorne and M. H. Wolpoff,
{\it The multiregional evolution of humans},
Scientific American {\bf 13}, (2003), 46-53

\end{thebibliography}
\end{document}